\documentclass[prl,twocolumn]{revtex4}
\usepackage{amssymb}
\usepackage{graphicx}

\def\v#1{\textbf{\emph{#1}}}
\def\c#1{\mathbb{#1}}

\def\braket#1{\left(#1\right)}
\def\sb#1{\left[#1\right]}

\def\ket#1{\left|#1\right\rangle}

\def\eps{\epsilon}
\def\up{\uparrow}
\def\down{\downarrow}

\def\skiip#1{}

\begin{document}
\title{Tensor-entanglement renormalization group approach to 2D quantum systems}
\author{Zheng-Cheng Gu$^\dagger$, Michael Levin$^{\dagger\dagger}$ and  Xiao-Gang Wen$^{\dagger}$}
\affiliation{ Department of Physics, Massachusetts Institute of
Technology, Cambridge, Massachusetts 02139, USA$^{\dagger}$
\\Department of Physics, Harvard University, Cambridge,
Massachusetts 02138, USA $^{\dagger\dagger}$ }

\begin{abstract}
Traditional mean-field theory is a simple generic approach for
understanding various phases. But that approach only applies to
symmetry breaking states with short-range entanglement. In this
paper, we describe a generic approach for studying
2D quantum phases with long-range entanglement (such as topological
phases). Our approach is a variational method that uses tensor product
states (also known as projected entangled pair states) as trial wave
functions. We use a 2D real space RG algorithm to evaluate expectation
values in these wave functions. We demonstrate our algorithm by studying
several simple 2D quantum spin models.

\end{abstract}
\maketitle

\emph{Introduction:}
To obtain various possible quantum phases of a quantum spin system $
H=\sum_{\<ij\>} \v S_i \cdot \v J_{ij} \cdot \v S_j$, we may use a mean-field
approach.  The mean-field approach can be viewed as a variational approach.
For example, to study the possible spin ordered phase of the above quantum
spin system, we may start with a trial wave function $ |\Psi_{trial}\>=\otimes
(u_i \mid\up\>_i +v_i \mid\down\>_i)$, where $\mid\up\>_i$ and $\mid\down\>_i$
are spin states on site $i$.  The spin ordered phases can be obtained through
minimizing average energy by changing $u_i$ and $v_i$. But such kind of
mean-field theory only apply to states with short-range entanglement (since
$|\Psi_{trial}\>$ is a direct product state).  As a result, we cannot use the
traditional mean-field theory to understand quantum phases that have pattern
of long-range entanglement (such as topologically ordered states and other
quantum states beyond Landau's symmetry breaking
description).\cite{W0275,KP0604,LWtopent}



\begin{figure}
\begin{center}
\includegraphics[width=2.7in]{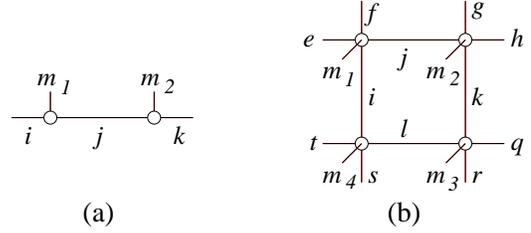}
\end{center}
\caption{Tensor-network -- a graphic representation of the
tensor-product wave function \eq{TPS}, (a) on a 1D chain  or (b) on
a 2D square lattice.  The indices on the links are summed over. }
\label{tps}
\end{figure}

One approach for addressing these phases is to use a more general
class of trial wave functions known as ``tensor product
states" (TPS) or ``projected entangled pair states" (PEPS).
\cite{FrankPEPS1,FrankPEPS2} Tensor product states were first discovered
in the context of the (1D) density matrix renormalization group (DMRG)
method \cite{WhiteDMRG,MPS}, but were later generalized to higher
dimensions and arbitrary lattices. On the square lattice (Fig. \ref{tps}),
the TPS are defined by
\begin{eqnarray}
\Psi(\{ m_i\})=\sum_{ijkl\cdots}
T_{ejfi}^{m_1}T_{jhgk}^{m_2}T_{lqkr}^{m_3}T_{tlis}^{m_4}\cdots
\label{TPS}
\end{eqnarray}
where $T_{ejfi}^{m_1}$ is a complex tensor with one physical index $m_i$
and four inner indices $i,j,k,l,\cdots$. The physical index runs
over the number of physical states $d$ on each site and inner
indices runs over $D$ values. Unlike simple mean field states,
these variational wave functions can describe 2D many-body quantum
systems\cite{FrankIsing} with short-range entanglement (such as symmetry
breaking states) as well as long-range  entanglement (such as string-net
condensed states\cite{GuString}).

One of main challenges of using this approach in higher dimensions
is that it is not easy to compute expectation values in these
states. In this Letter, we describe a simple solution to this
problem in two dimensions. Our approach - which we call the tensor
entanglement renormalization group (TERG) method - is an
approximation scheme based on the 2D real space RG method developed
in \Ref{LevinTRG}. A different real space RG method can be found in
Ref. \cite{VidalRG}.

As we mentioned earlier, this kind of variational approach has the
advantage that it can potentially address 2D quantum many-body states
that contain \emph{both} symmetry breaking orders and topological orders. In
this paper, we will just introduce our algorithm by studying a few simple
quantum models and compare our results with
those obtained through other previous methods. The application of TERG
approach to topologically ordered states will be presented in future
publications.

To see the efficiency of the TERG method, let us compare it with
other variational methods for 2D gapped systems (see Appendix for an
explanation):
\begin{center}
\begin{tabular}{|l|l|} \hline
Method & Error\\
\hline
VQMC  & $\eps \sim 1/T^{1/2}$\\
1D approach & $\epsilon \sim \exp(-\text{const} \cdot \log T)$\\
TERG   &  $\eps \sim \exp(-\text{const} \cdot (\log T)^2)$\\
\hline
\end{tabular}
\end{center}
Here $T$ the calculation time and $\eps$ is the achieved accuracy of
the calculated average energy for a given many-body state. The acronym
VQMC stands for variational Quantum Monte Carlo, while ``1D approach"
refers to an approximation scheme where one replaces the infinite 2D
lattice by an $L \times \infty$ lattice, for $L$ large but finite, and
then computes expectation values using a transfer matrix approach.

\begin{figure}
\begin{center}
\includegraphics[width=3.5in]{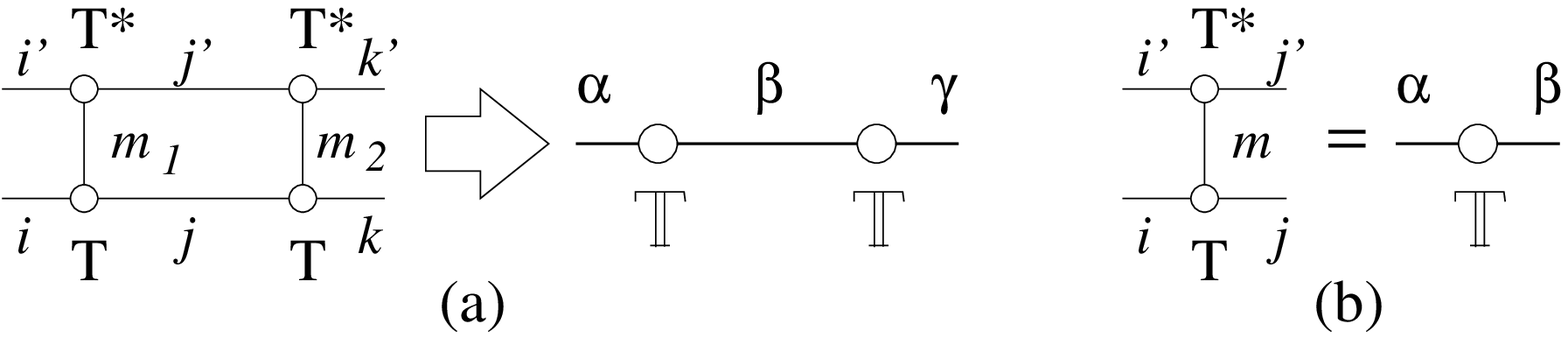}
\end{center}
\caption{ (a) The graphic representation of the inner product of
TPS, $\<\Psi|\Psi\>$, in term of the tensor $T$ or the double-tensor
$\c T$ (see \eq{avaH}).  (b) After summing over $m$ and identifying
$(i,i') \to \al$, $(j,j') \to \bt$, we obtain the double-tensor $\c
T$ from $T$ and $T^*$. } \label{Str}
\end{figure}

\begin{figure}
\begin{center}
\includegraphics[width=3.0in]{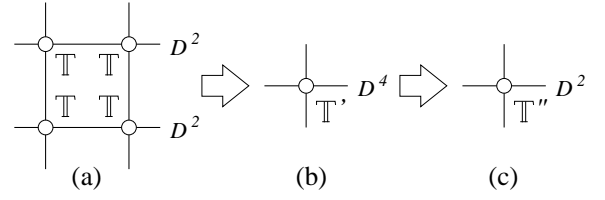}
\end{center}
\caption{ The indices of the double-tensor have a range $D^2$. After
combine the two legs on each side into a single leg, the four linked
double-tensors in (a) can be viewed as a single double-tensor $\c
T'$ whose indices have a range $D^4$. (c) $\c T'$ can be
approximately reduced to a ``smaller'' double-tensor $\c
T^{\prime\prime}$ whose indices have a range $D^2$ and satisfies $
\text{tTr} [\c T^{\prime}\otimes \c T^{\prime} \cdots] \approx
\text{tTr} [\c T^{\prime\prime}\otimes \c T^{\prime\prime} \cdots]
$. } \label{RG}
\end{figure}

\emph{The TERG method:}  Let us consider a system with
translationally invariant $H=\sum_i H_i$.  $H_i$ can always be
expressed as a summation of products of local operators:
$H_i=\hat{O}_i^0+\hat{O}_i^1 \hat{O}_{j}^2 + \cdots$. A key step in
the variational approach is to calculate the norm and the
expectation value of $H_i$ for the trial wave function $\Psi(\{
m_i\})$:
\begin{eqnarray}
\label{avaH} \langle \Psi |\Psi\rangle &=& \sum_{m_1m_2\cdots}
\sum_{ii'jj'\cdots} T_{ejfi}^{m_1} T_{e'j'f'i'}^{m_1*}
T_{jhgk}^{m_2} T_{j'h'g'k'}^{m_2*} \cdots
\nonumber\\
&=&
{\rm{tTr}} \sb{\c T\otimes \c T\otimes\c T\otimes   \cdots},
\nonumber\\
\langle \Psi |H_i|\Psi\rangle &=& {\rm{tTr}} \sb{\c T^0_i\otimes \c
T\otimes\c T\otimes   \cdots}
\nonumber\\
&+&{\rm{tTr}} \sb{
\c T^1_i\otimes \c T_{j}^2 \otimes\c T\otimes
\cdots} +\cdots
\end{eqnarray}
where double-tensors $\c T$, $\c T^a$, $a=0,1,2$, are defined as
(see Fig. \ref{Str}b)
\begin{eqnarray}
\c T= \sum_{m} {T^{m*}} \otimes T^{m};\quad \c T^a=\sum_{mm^\prime}
O^a_{mm^\prime} {T^{m*}} \otimes T^{{m}}
\end{eqnarray}
with $O_{mm^\prime}^a$ the matrix elements of local operators
$\hat{O}^a$ in the local basis $\ket{m}$. The tensor-trace (tTr)
here means summing over all indices on the connected links of
tensor-network (see Fig. \ref{RG}a). Note that the inner product is
obtained from a uniform tensor-network.  The average of the on-site
interaction is obtained from a tensor-network with one ``impurity''
tensor $\c T^0$ at site $i$ (while other site has $\c T$).
Similarly, for two-body interactions, the tensor-network has two
``impurity'' tensors $\c T^1$ and $\c T^2$ at $i$ and $j$.

Calculating the tensor-trace $\text{tTr}$ is an exponentially hard calculation
in 2D or higher dimensions.  Motivated by the tensor-renormalization approach
developed in Ref. \cite{LevinTRG}, we can accelerate the calculation
exponentially if we are willing to make an approximation. The basic idea is
quite simple and is illustrated in Fig. \ref{RG}. After finding the reduced
double-tensor $\c T^{\prime\prime}$, we can express $\text{tTr}[\c T\otimes \c
T\cdots] \approx \text{tTr}[\c T^{\prime\prime}\otimes \c
T^{\prime\prime}\cdots ]$ where the second tensor-trace only contain a quarter
of the double-tensors in the first tensor-trace.  We may repeat the procedure
until there are only a few double-tensor in the tensor-trace. This allows us
to reduce the exponential long calculation to a polynomial long calculation.

\begin{figure}
\begin{center}
\includegraphics[width=3.5in]
{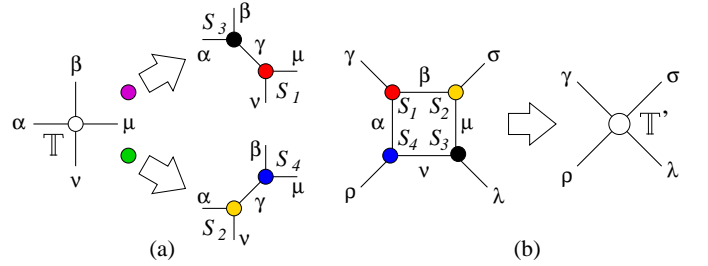}
\end{center}
\caption{(Color online) (a) We represent the original rank-four tensor by two
rank-three tensors, which is an \emph{approximate} decomposition.
(b) Summing over the indices around the square produces a single tensor
$\c T'$.  This step is \emph{exact}.
} \label{tsrd}
\end{figure}

The detail implementation of the above TERG approach is actually a
little more involved.  For an uniform tensor-network Fig \ref{RG}a, we
can coarse grain it in two steps. The first step is decomposing the
rank-four tensor into two rank-three tensors. We do it in two
different ways on the sublattice purple and green (see Fig.
\ref{tsrd}a). On purple sublattice, we have $\c T_{\al \bt \mu
\nu}=\sum_{\ga'} {S_1}_{\mu\nu \ga'}{S_3}_{\al\bt \ga'}$ and on
green sublattice, we have $\c T_{\al \bt \mu \nu}=\sum_{\ga'}
{S_2}_{\nu\al\ga'}{S_4}_{\bt\mu\ga'}$.  Note that $\al,\bt,\mu,\nu$
run over $D^2$ values while $\ga'$ run over $D^4$ values.

Next we try to reduce the range of $\ga'$ through an
approximation.\cite{LevinTRG} Say, on purple sublattice, we view $\c
T_{\al\bt\mu\nu}$ as a matrix $M_{\al\bt;\mu\nu}^{\rm{red}}=\c
T_{\al\bt\mu\nu}$ and do singular value decomposition $M^{\rm{red}}=U\La
V^\dagger$.  We then keep only the largest $D_{cut}$ singular values $\la_\ga$
and define ${S_1}_{\mu\nu\ga}=\sqrt{\la_\ga}V^\dag_{\ga,\mu\nu}$,
${S_3}_{\al\bt\ga} =\sqrt{\la_\ga}U_{\al\bt,\ga}$.  Thus, we can approximately
express $\c T_{\al\bt\mu\nu}$ by two rank-three tensors $S_1$, $S_3$
\begin{eqnarray}
\c T_{\al\bt\mu\nu}\simeq\sum_{\ga=1}^{D_{\rm{cut}}}
{S_3}_{\al\bt\ga} {S_1}_{\mu\nu\ga}
.
\label{rule1}
\end{eqnarray}
Similarly, on green sublattice we may also define $\c T_{\al\bt\mu\nu}$ as a
matrix $M_{\nu\al;\bt\mu}^{\rm{green}}$ and do singular value
decompositions, keep the largest $D_{\text{cut}}$ singular values
and approximately express $\c T_{\al\bt\mu\nu}$ by two rank three tensors
$S_2,S_4$.
\begin{eqnarray}
\c T_{\al\bt\mu\nu}\simeq\sum_{\ga=1}^{D_{\rm{cut}}}
{S_2}_{\nu\al\ga}{S_4}_{\bt\mu\ga}\label{rule2}
\end{eqnarray}

\begin{figure}
\begin{center}
\includegraphics[width=2.5in]
{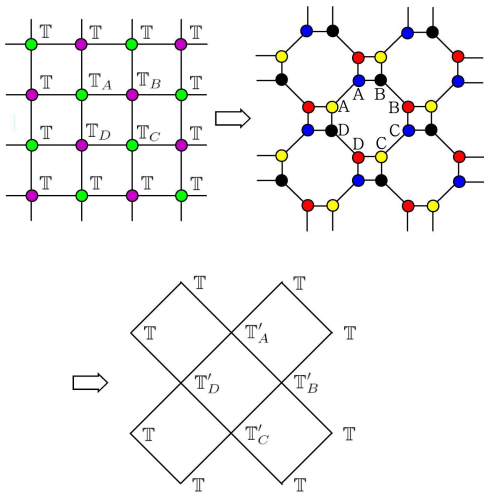}
\end{center}
\caption{(Color online) Iteration of tensor-network produces a coarse grained tensor-network.}
\label{energy1}
\end{figure}

After such decompositions, the square lattice is deformed into the form in
Fig. \ref{tsrd}b (see also Fig. \ref{energy1}).  The second step is simply
contract the square and get a new tensor on the coarse grained lattice.
\begin{eqnarray}
\c T_{\ga\si\la\rho}^\prime=\sum_{\al\bt\mu\nu} {S_1}_{\bt\al\ga}
{S_2}_{\mu\bt\si} {S_3}_{\nu\mu\la} {S_4}_{\al\nu\rho}
\end{eqnarray}
The range of indices for the reduced double-tensor $\c T'$ is only
$D_\text{cut}$ which can be chosen to be $D^2$ or some other values.
Repeat the above two steps twice, we can get the reduction from Fig.
\ref{RG}a to \ref{RG}c.

The above TERG procedure can be easily generalized to tensor-network
with ``impurities'', such as the one in Fig. \ref{energy1} which has
four ``impurity'' tensors. Evaluating Fig. \ref{energy1} will allow
us to calculate the averages of up to four-body nearest-neighbor
interactions (which include on-site interaction, nearest-neighbor
and next-nearest-neighbor two-body interactions).  TERG procedure is
illustrated in Fig. \ref{energy1}. We note that the number and the
relative positions of ``impurity'' tensors do not change after each
iteration.  So we repeat the same iterative calculation until there
are only a few tensors in the tensor-trace.  Thus the calculations
of the averages of local operators is also reduced to polynomial
long calculations.  The total computational complexity is
$\text{cost time}\sim  D_{\text{cut}}^6 \text{log} N$ on square
lattice ($N$ is the total number of sites).  For gapped systems in
the thermodynamic limit, the truncation error can be estimated as
$\epsilon \sim \exp \sb{-\text{const} \cdot \braket{\log
D_{\text{cut}}}^2}$\cite{LevinTRG}. After calculating the inner
product and the average of $H_i$ in \eq{avaH}, we can obtain the
approximated ground state with minimized average energy by adjusting
the elements in the tensor $T$.

\emph{Examples:}
To test our TERG algorithm, we first calculate ground state and its
magnetization along $x$ and $z$ directions for the transverse field Ising
model:
\begin{eqnarray}
\label{tIsingeq}
H=-\sum_{\langle ij \rangle}\sigma_i^z\sigma_j^z-h\sum_i \sigma^x_i
\end{eqnarray}
We choose the tensor $T$ in Eq. (\ref{TPS}) to be real and has 90 degree
rotational symmetry. We also choose the inner dimension $D=2$ and keep 18
singular values at each iteration ($D_{\text{cut}}=18$).  The total system
size is up to $2^9\times 2^9$ sites.  The average energy for a tensor $T$ is
calculated using the TERG approach. We use Powell minimization method to find
the minimal average energy and the corresponding tensor which gives us the
variational ground state.

\skiip{
\begin{figure}
\begin{center}
\includegraphics[width=2.8in]
{sigularvalue.eps}
\end{center}
\caption{(Color online) The log plot of singular value distributions at each
step, both for non-critical system (a) $h=2$ and critical system (b)
$h=h_c=3.08$.
} \label{singularvalue}
\end{figure}
In Fig.\ref{singularvalue}, we plot the singular values distributions at
several coarse-grained steps when calculating the norm
$\langle\Psi|\Psi\rangle$(Because the symmetry of our ansatz, the singular
values are the same for sublattice A and B. The largest singular value is
normalized to 1 at each step.) For first and second steps($n=1,n=2$), there
are only 16 singular values, so our results are \emph{exact}. From $n=3$, we
always keep the largest $D_{\text{cut}}$ singular values and after several
steps $n=5,n=6$, the sequence of singular values converges.  For non-critical
systems, the singular value distributions decay much faster than polynomial
and decay polynomially for critical systems. Thus, in practice many critical
properties can be accessed by considering large - but finite - system sizes.
The variational ground state is obtained by minimizing the average energy by
choosing a proper tensor.
}

\begin{figure}
\begin{center}
\includegraphics[width=3.6in] {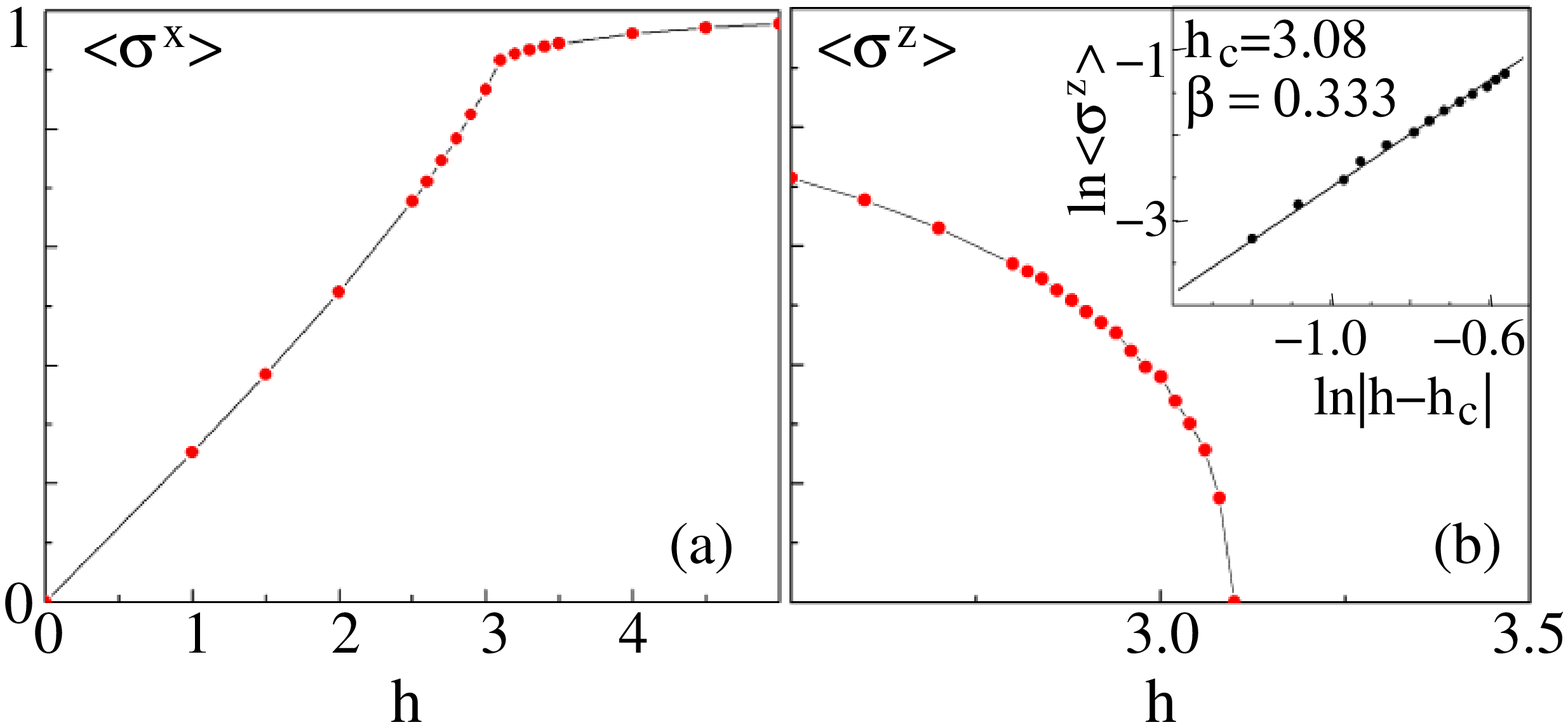}
\end{center}
\caption{(Color online)
(a)
Magnetization along the $x$ direction $\langle
\sigma^x\rangle$ versus transverse field $h$.
The derivative of magnetization has a singularity
around $h\simeq 3.1$, indicating
the second order phase transition.
(b) Magnetization along the $z$ direction $\langle
\sigma^z\rangle$ versus transverse field $h$. In the inset is the
log plot of $\langle \sigma^z\rangle$ versus $|h-h_c|$, where $h_c$
is the critical field.
} \label{tIsing}
\end{figure}

In Fig. \ref{tIsing} we plot the polarization along $x$
direction and $z$ direction in the variational ground state.  We note that
despite the $\si^z \to -\si^z$ symmetry in the Hamiltonian, the tensor $T$
that minimize the average energy may break the $\si^z \to -\si^z$ symmetry and
give rise to non-zero polarization in $z$ direction.  We find a second order
phase transition at $h_c\approx3.08$. We further fit the critical exponent
\begin{eqnarray}
\langle \sigma^z\rangle=A|h-h_c|^\beta
\end{eqnarray}
with $\beta\approx0.333\pm 0.003$.  Both the values of critical
field and critical exponent $\beta$ here are very close to the QMC
results, with $h^{QMC}_c\simeq 3.044$\cite{QMC1} and
$\beta^{QMC}\simeq 0.327$\cite{QMC2}. They are much better than the
meanfield results $h_c=4$ and $\beta=0.5$.

\begin{figure}
\begin{center}
\includegraphics[width=2.0in] {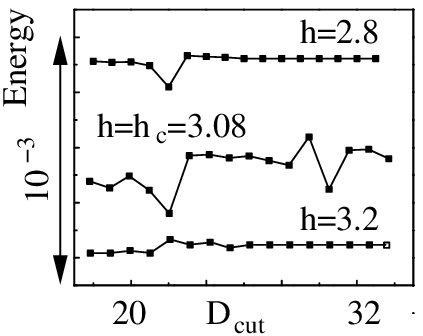}
\end{center}
\caption{
Ground state energies of the transverse Ising model for different $D_{cut}$.
}
\label{error}
\end{figure}

To see the truncation error caused by $D_{cut}$, we plot the ground
state energy (per site) of Eq. \ref{tIsingeq} as a function of
$D_{cut}$, for $h=2.8,3.2$, and $h=h_c=3.08$ (see Fig. \ref{error}).
The energies for different $h$'s are shifted by different constants
so that the three curves can be fitted into one window.  Notice that
for off critical systems ($h=2.8,3.2$), the energy converges very
quickly for small $D_{cut}$ ($\sim 26$).  Even at the critical point
$h=h_c=3.08$, the error in energy per site is of order $10^{-4}$.
The truncation error is much smaller for gapped off-critical states.

As another more stringent test, we also apply the TERG method to
study Heisenberg model $H=\sum_{\<ij\>} \v S_i \cdot \v S_j$ on
square lattice which contains gapless excitations.  Again we choose
$D=2$, $D_{cut}=18$ and total system size $2^9\times 2^9$ sites.  We
choose tensors $T_A$ and $T_B$ to be real and has 90 degree
rotational symmetry on sublattice $A$ and $B$. We find the ground
state energy to be $-0.33$ per bond, which is quite close to the
best QMC results(-0.3350).\cite{SandvikAF}  The TERG method also
allows us to calculate correlation function using tensor-network
with two ``impurity'' tensors with arbitrary separations.
Through the long-range correlation function, we find that the total
magnetization is $m=\sqrt{\langle S_i^x S_j^x +S_i^y S_j^y +S_i^z
S_j^z \rangle}=0.39$, which is larger than the QMC
results(0.307).\cite{SandvikAF} We see  that a small error in ground
state energy (which depends only on short-range correlation) can
leads to a larger error on correlations at long distances.

\emph{Conclusions and discussions:}
The TERG approach is a simple generic method to obtain various quantum
phases and quantum phase transitions for quantum systems in any
dimension. The most important feature of TERG approach is that it can
handle quantum states with long-range entanglement (such as topologically
ordered states).  When we use traditional mean-field theory to calculate
quantum phase diagram, the topological ordered phases cannot appear in
such a mean-field phase diagram, since the mean-field states are limited
to those with short-range entanglement. The TERG approach solves this
problem and can generate phase diagrams that contain both symmetry
breaking states and topologically ordered states.

\emph{Acknowledgements:}
We would like to thank Frank Verstraete for very helpful discussions and
comments.  This research is supported by the Foundational Questions Institute
(FQXi) and NSF Grant DMR-0706078.

\emph{Appendix:} In a VQMC calculation, the error $\eps$ is a statistical
error that scales like $1/N^{1/2}$ where $N$ is the number of samples. The
computational time $T$ scales like $N$. Thus the scaling of the error with
computational time is given by $\eps \sim 1/T^{1/2}$.

In the 1D approach, the error $\eps$ is a finite size error that comes
from the truncation of the infinite 2D lattice to an $L \times \infty$
lattice. In a gapped system we expect this error to fall off as
$e^{-L/\xi}$ where $\xi$ is the correlational length.
On the other hand, the computational time $T$ is exponential in $L$ since
the method requires diagonalizing a transfer matrix whose size is
exponentially large in $L$. We conclude that the error scales with
computational time as $\eps \sim e^{-\text{const.} \cdot \log T}$.

In the TERG approach, the truncation error for each iteration step
scales as $\eps_1 \sim  e^{-\text{const.} \cdot (\log D_{\text{cut}})^2}$,
since calculating the norm and averages is like calculating the partition
function in \Ref{LevinTRG}. The total truncation error for a system of
size $L$ is $\eps_t \sim (\log L) e^{-\text{const.} \cdot (\log
D_{\text{cut}})^2}$
since such a system requires $\log L$ iterations. On the other hand, the
finite size error is $\eps_s\sim e^{-L/\xi}$. Minimizing the sum of
the two errors, we see that the optimal $L$ is given by
$L \sim (\log D_{\text{cut}})^2$. Since the computational time scales
polynomially in $D_{\text{cut}}$, we conclude that the total error scales
like $\eps \sim e^{-\text{const.} \cdot (\log T)^2}$ (neglecting
subleading log corrections).


\end{document}